\documentclass[12pt]{iopart}
\usepackage{iopams,epsfig}
\begin{document}
\jl{3}

\title[Energy landscapes and High-field ESR]{
A study of the deep structure of the energy landscape of glassy polystyrene: 
the exponential distribution of the energy-barriers revealed by high-field 
Electron Spin Resonance spectroscopy}

\author{V. Bercu\dag \ddag \S, M.Martinelli\dag ,C.A.Massa\dag, 
L.A.Pardi\dag , D~Leporini\ddag \S $\|$ \footnote[5]{To
whom correspondence should be addressed.}}

\address{\dag IPCF-CNR, I-56100 Pisa, Italy}

\address{\ddag Dipartimento di Fisica ``Enrico Fermi'',
Universit\`a di Pisa, via F.\@ Buonarroti 2, I-56127 Pisa, Italy}

\address{\S  INFM UdR Pisa ,via F.\@ Buonarroti 2, I-56127 Pisa, 
Italy}

\address{$\|$ INFM-CRS SOFT , 
Universit\`a di Roma La Sapienza, P.zza A. Moro 2, I-00185, Roma, 
Italy}

\begin{abstract}
The reorientation of one small paramagnetic molecule ( spin probe ) 
in glassy polystyrene ( PS ) is studied by high-field Electron Spin 
Resonance spectroscopy at two different Larmor frequencies ( $190$ and 
$285 \; GHz$ ). The exponential distribution of the energy-barriers  
for the rotational motion of the spin probe is unambigously evidenced 
at both $240\;K$ and $270\;K$.  The same shape for the distribution of the 
energy-barriers of PS was evidenced by the master curves 
provided by previous mechanical and light scattering studies.
The breadth of the energy-barriers distribution of the spin probe is 
in the range of the estimates of the breadth of the PS energy-barriers
distribution. 
{ \bf The evidence that the deep structure of the energy landscape of PS
exhibits the exponential shape of the energy-barriers distribution agrees 
with results from extreme-value statistics and the trap 
model by Bouchaud and coworkers }. 

\end{abstract}

\submitto{\JPCM}
\pacs{64.70.Pf, 76.30.-v, 61.25.Hq}


\ead{dino.leporini@df.unipi.it}
\maketitle

\section{Introduction}
\label{introduction}
The study of glassy solid dynamics is a very active one 
\cite{AngEtAl00,review}.
Here one is interested in the temperature range which is, on the one 
hand, well below the glass transition temperature $T_{g}$ to neglect 
aging effect and consider the glassy system as one with constant 
structure and, on the other hand, high enough to neglect tunneling 
effects  governing the low-temperature anomalies of glasses. 
In this regime the dynamics is thermally activated in the 
substructures of the minima of the energy landscape accounting for 
various subtle degrees of freedom \cite{Angell98}. It has been shown that in the 
glass the 
temperature dependence of the energy barrier distribution $g(E)$ is 
only weakly temperature-dependent \cite{Wu90,WuNagel90}.

The shape of the energy-barriers distribution $g(E)$ in glasses 
has been extensively investigated via experiments 
\cite{review,Wu90,WuNagel90,qi,rosslervogel1,rosslervogel2,sokolov0,sokolov,
Topp,GorhametAl2004}, theories \cite{Derrida81,Scher,Bassler87,Odagaki95,
Bouchaud1,Bouchaud2,RinnEtAl2001,SchonSibani2000} and 
simulations \cite{Bouchaud3}. Basically, two different distributions 
are usually recovered, the gaussian distribution
\cite{review,Wu90,WuNagel90,qi,rosslervogel1,sokolov0,
Derrida81,Bassler87,Bouchaud3} and the exponential distribution 
\cite{sokolov0,sokolov,Topp,Scher,Odagaki95,Bouchaud1,Bouchaud2,RinnEtAl2001,
SchonSibani2000}. 
The convolution of these two distributions \cite{RosslerEtAl90a} 
as well as the truncated Levy fligth, i.e. a power law with exponential cutoff, 
resembling the stretched exponential \cite{GorhametAl2004} were also considered.

It is interesting to relate $g(E)$ with the density of states, i.e. 
the distribution of the minima of the energy landscape. 
On the upper part of the landscape, being explored at high temperatures, 
the Central Limit theorem suggests that the density is gaussian 
\cite{Derrida81,Bouchaud3}. At lower temperatures the state point is trapped 
in the deepest low-energy states which
are expected to be exponentially distributed following general 
arguments on extreme-value statistics leading to the so-called 
Gumbel distribution \cite{Bouchaud2}. Random Energy models \cite{Derrida81} 
and numerical simulations \cite{SchonSibani2000} support the conclusion.
The barrier height $E$ during the jump from one state with energy $E_{1}$ to 
another state with energy $E_{2}$ has been modelled by the linear combination 
$E = \alpha E_{2} + (1- \alpha )E_{1}$ \cite{RinnEtAl2001}.  In the case 
of trap models ( $\alpha =0$ ) the minima and the energy barriers have 
the same distribution. This also holds true for 
annealed disorder and $\alpha > 0$ \cite{Bouchaud1,RinnEtAl2001}. 
Moreover, one notes that the linear combination of independent gaussian 
variables is gaussian too and the linear combination of independent 
exponential variables has exponential tail. All in all strict 
relations are expected between $g(E)$ and the density of states.

If the average trapping time $\tau$ before to overcome the 
barrier of height $E$  at temperature $T$ is governed 
by the Arrhenius law, 

\begin{equation}
    \tau = \tau_{0} exp(E/kT)
    \label{eq:Arrhenius}
\end{equation}

\noindent
$k$ being the Boltzmann's constant, the distribution of barrier heights 
induces a distribution of trapping times $\rho (\tau )$. The explicit 
form of $\rho (\tau )$ for a gaussian distribution of barrier heights 
with width $\sigma_{E}$ is the log-gauss distribution ( LGD )

\begin{equation}
\rho_{LGD} (\tau) =
\frac{1}{\sqrt{2\pi \sigma ^{2}}}\exp \left[ -\frac{1}{2\sigma 
^{2}}\left(
\ln \frac{\tau }{\tau_{LGD}}\right) ^{2}\right] \frac{1}{\tau}
       \label{eq:distrloggauss}
\end{equation}

\noindent
$\sigma = \sigma_{E}/kT $ is the width parameter. $ \rho (\tau )$ 
is maximum at $\tau_{LGD}$. If the distribution of barrier heights is 
exponential with average height $\overline{E}$

\begin{equation}
    g(E) =  \frac{1}{\overline{E}} exp( - \frac{E}{\overline{E}})
    \label{eq:energydistribution}
\end{equation}

\noindent
$\rho (\tau )$ is expressed by the power-law distribution ( PD )

\begin{equation}
\rho_{PD} (\tau ) = \left \{ 
\begin{array}{ll}
0                           &\mbox{if $\tau <   \tau_{PD}$} \\
x \tau _{PD}^{x}\tau ^{-(x+1)}&\mbox{if $\tau \ge \tau_{PD}$}
\end{array} \right.
       \label{eq:distrpower}
\end{equation}

\noindent
with $x= kT/ \overline{E}$ and $\tau_{PD} =\tau_{0} $. If the width of 
the energy-barriers distribution is vanishingly small, a single trapping 
( correlation ) time ( SCT ) is found and both $\rho_{LGD}$ and $\rho_{PD} $ 
reduce to:

\begin{equation}
    \rho_{SCT} (\tau ) = \delta(\tau-\tau_{SCT})
    \label{eq:deltadistr}
\end{equation}

The shape of $g(E)$ has been usually studied by measuring 
different susceptibilities $\chi''(\nu)$ arising from either 
collective 
or single-particle response. If the susceptibility
follows by a static distribution of activated relaxation times in the 
presence of a wide distribution $g(E)$ it follows \cite{Bordewijk}

\begin{equation}
    \chi''(\nu) \propto T g(E) , \hspace{1cm} E = kT ln(1/2 \pi \nu 
\tau_{0})
    \label{eq:scaling}
\end{equation}

\noindent
The above equation shows that the shape of $\chi''(\nu)$ yields the 
shape of $g(E)$. The conversion factor between 
the frequency and the energy scales is the unknown attempt frequency 
$1/\tau_{0}$ which is usually treated as one adjustable  parameter to 
set both the width and the location of $g(E)$. The dielectric 
spectroscopy provides a convenient frequency range to recover the 
full shape of $\chi''(\nu)$ and then of $g(E)$ \cite{Wu90,WuNagel90}. 
In other cases, e.g. light scattering \cite{sokolov} and mechanical 
relaxation \cite{Topp}, the accessible frequency range is limited and 
$g(E)$ is recovered by building suitable master curves assuming the 
time-temperature superposition principle. Nuclear Magnetic Resonance 
( NMR ) offers an alternative procedure to get information on $g(E)$ 
by analyzing the NMR lineshape in terms of two weighted components 
\cite{review,RosslerEtAl90b}. However, an adjustable conversion factor 
between the frequency and the energy scales is also needed.

The use of suitable probes to investigate the secondary relaxations 
in glasses by NMR \cite{review,qi,RosslerEtAl90a,RosslerEtAl90b},
Electron Spin Resonance ( ESR ) 
\cite{AndrEtAl96,FaetEtAl99,BarbEtAl04} and 
Phosphorescence \cite{ChristoffAtvars99} studies is well documented. 
In spite of that efforts, the relation between the probe motion and 
the host dynamics is usually not obvious with few exceptions 
which,notably, involved relatively small and nearly spherical 
probe molecules \cite{FaetEtAl99}. It was also noted that 
small molecules, e.g. xanthone and benzophenone, are more sensitive 
to shorter segmental motions occurring at lower temperatures 
\cite{ChristoffAtvars99}.

During the last few years continuous-wave ( CW ) and pulsed High-Field ESR 
(~HF-ESR ) techniques were developed involving large polarizing magnetic 
fields, e.g. $B_{0} \cong 3 T$ corresponding to Larmor frequencies about $95 
GHz$ ( W band ), \cite{lebedev,freed2,esrgunnar}. 
HF-ESR is widely used in solid-state physics \cite{MartinelliEtAl2003}, biology 
\cite{LiangFreed99a,LiangEtAl99b,BorbatEtAl2001,FuchsEtAl2002,SchaferEtAl2003}  
and polymer science 
\cite{PilarEtAl00,LepEtAl02a,LepEtAl02b,LepEtAl03}.
One major feature is the remarkable orientation resolution 
\cite{LepEtAl03}
due to increased magnitude of the anisotropic Zeeman interaction 
leading to a wider distribution of resonance frequencies 
\cite{berliner,Muus}. 

In the present paper one demonstrates that the rotational motion of 
suitable small guest molecules ( spin probes ),  as detected by 
HF-ESR , is an effective probe sensing the energy-barrier 
distribution $g(E)$ of glassy polystyrene ( PS ). The choice of PS was 
motivated by a number of studies of $g(E)$ by Raman \cite{sokolov0} 
and light scattering \cite{sokolov} 
and mechanical relaxation \cite{Topp} which evidenced its exponential form 
( eq.\ref{eq:energydistribution} ). The distribution of barriers to 
be surmounted by suitable probes in PS was also studied by NMR 
\cite{RosslerEtAl90a}. The secondary relaxations of PS were 
investigated by phosphorescent probes \cite{ChristoffAtvars99}.

The paper is organized as follows. In Sec. \ref{background} the 
background on ESR is presented. Experimental details are given in 
Sec.\ref{Experimental} and the results are discussed in Sec. \ref{Results}.
The conclusions are summarized in Sec. \ref{conclusions}.

\section{ESR Background}
\label{background}

The main broadening mechanism of the ESR lineshape is the coupling 
between the 
reorientation of the spin probe and the relaxation of the electron 
magnetisation
\textbf{M} via the anisotropy of the Zeeman and the hyperfine 
magnetic 
interactions. When the molecule rotates, the coupling gives rise to 
fluctuating 
magnetic fields acting on the spin system. The resulting phase shifts 
and 
transitions relax the magnetisation and broaden the resonance 
\cite{berliner,Muus}. 
Because of the roughness of the energy landscape and the highly 
branched
character of the free volume distribution, one expects that the 
small spin probe undergoes jump dynamics in glasses . 
One efficient numerical approach to calculate
the ESR line shape in the presence of rotational jumps is 
detailed elsewhere \cite{LepEtAl02a}.

The rotational correlation time $\tau_{c}$, i.e. the area below the
correlation function of the spherical harmonic Y$_{2,0}$, is 
\cite{AndrEtAl96}

\begin{equation}
\tau_{c} =\frac{\tau ^{\ast }}{\left[ 1-\frac{\sin (\frac{5\phi 
}{2})}{5\sin (\frac{
\phi }{2})}\right] }
       \label{eq:tau}
\end{equation}

\noindent
where $\phi $ and $\tau ^{\ast }$ are the size of the angular jump and
the mean residence time before a jump takes place, respectively. 
In the limit $\phi <<1$ the jump model reduces to the isotropic 
diffusion model with $\tau_{c} = 1/6D = \tau ^{\ast} /\phi ^{2}$
where $D$ is the rotational diffusion coefficient.

The occurrence of a static distribution of correlation times in glasses 
suggests to evaluate the ESR line shape $L(B_{0})$, which is usually 
detected by sweeping the static magnetic field $B_{0}$ and diplaying 
the first derivative, as a weighted superposition of different contributions:

\begin{equation}
L(B_{0})=\int_{-\infty }^{+\infty }
d\tau_{c} L(B_{0},\tau_{c} )\rho (\tau_{c} )
\label{eq:superpos}
\end{equation}

\noindent
$L(B_{0},\tau_{c} )$ is the EPR line shape of the spin probes with 
correlation time $\tau_{c} $ and $\rho (\tau_{c} )$ is the $\tau_{c} $ 
distribution.

It must be pointed out that in the presence of wide distribution of 
correlation times the ESR lineshape ( eq. \ref{eq:superpos} ) cannot be 
simplified as a sum of few leading components. This is rather 
different with respect to NMR  where `two-phase spectra' stemming 
from 
`fast' and `slow' molecules  are observed \cite{review}.
If, on one hand, that feature leads to more elaborated numerical 
work, on the other hand it ensures that the ESR lineshape exhibits a much 
richer variety of details being markedly affected by the spin-probe 
dynamics which provides helpful constraints to the fit procedures and 
results in an improved information content of the spectra.

\section{Experimental}
\label{Experimental}

PS was obtained from Aldrich and used as received.  The 
weight-average molecular 
weight is $M_{w}=230 kg/mol$ and $T_{g}=367K$. The free radical used 
as spin 
probe was 2,2,6,6-Tetramethyl-1-piperidinyloxy (TEMPO) from Aldrich. 
TEMPO
has one unpaired electron spin $S=1/2$ subject to hyperfine 
interaction with 
the nitrogen nucleus with spin $I=1$. The sample was prepared by the 
solution 
method \cite{SaalmuellerEtAl96} by dissolving TEMPO and PS in 
chloroform. 
The solution was transferred on the surface of one glass slide and 
heated at 
$T_{g}+10K$ for 24h. The spin probe concentration was less than 
0.08\% in weight. 
Appreciable broadening  of the ESR line shape due to the spin-spin 
interaction 
are observed for concentration larger than\ 0.2\% in weight. No 
segregation of 
the spin probe was evidenced. Samples aged at room temperature for 
six months 
exhibited no appreciable changes of the ESR lineshape.

The ESR experiments were carried out on the ultrawide-band ESR 
spectrometer which is detailed elsewhere \cite{AnninoEtAl00}. In the 
present study the spectrometer was operated at two different 
frequencies, i.e. at $190 GHz$ and $285 GHz$. The multi-frequency 
approach ensures better accuracy to determine the spin-probe dynamics.  
The sample of about $0.8 cm^{3}$ was placed in a Teflon sample holder 
in a single-pass probe cell. All spectra were
recorded and stored in a computer for off-line analysis. 

\section{Results and Discussion}
\label{Results}

The numerical simulation of the ESR lineshape much relies on the 
accurate determination of the magnetic parameters of the spin probe. 
To this aim, one profited from the enhanced resolution of HF-ESR 
and measured the ESR lineshape of TEMPO in PS at low 
temperature where the rotational motion of the spin probe is very 
slow and the resulting lineshape approaches the so called `powder' or 
`rigid' limit lineshape \cite{Muus}. The results at $50 K$ are shown in 
Fig.\ref{FIG1}. It is seen that the lineshapes at both the operating 
frequencies are well fitted by the SCT model with a single set of magnetic 
parameters. The small discrepancy between the simulation and the lineshape 
at low magnetic field was already noted in other studies 
\cite{freed2}. Remarkably, no distribution of correlation times 
is evidenced at $50 K$. At this stage a complete discussion of this 
finding needs the detailed investigation of the temperature 
dependence of the rotational dynamics of TEMPO in PS. This is beyond 
the purpose of the present Letter and will be discussed thoroughly in 
a forthcoming paper \cite{jcpfuture}. 

Fig.\ref{FIG2} compares the ESR lineshapes at $190 GHz$ of TEMPO 
in PS at $T = 270K = T_{g}-97K$ with the SCT ( left, two adjustable 
parameters $\tau_{SCT},\phi$ ) and LGD models ( right, three adjustable 
parameters $\tau_{LGD},\phi$ and $\sigma$ ). For a given jump angle $\phi$
the remaining one ( SCT ) or two ( LGD ) parameters have been adjusted.
It is apparent that the SCT model is inadequate. Large disagreements 
were also found when fitting the ESR lineshape at $285 GHz$ ( not shown). 
Fig.\ref{FIG2} shows that the LGD model yields better agreement for large 
jump angles. However, closer inspection reveals deviations in 
the wings of the ESR lineshapes at both low- and high-field. The ESR 
absorption of these regions is mainly contributed by spin probes with 
their $x$ molecular axis (low-field) or $z$ axis ( high field) being parallel 
to the static magnetic field  \cite{LepEtAl03}. Being the magnetic 
parameters of the spin probe precisely measured, it may be shown 
that the above disagreements are clear signatures of the 
overestimate of the jump angle size \cite{jcpfuture}. 
The best-fit value of the width parameter $\sigma = 1$ of the LGD 
distribution (eq. \ref{eq:distrloggauss}) corresponds to the width 
$\sigma_{E}/k = 270 K $ of the energy-barrier distribution of TEMPO in PS. 
R\"{o}ssler and coworkers found by NMR $\sigma_{E}/k = 276 K $ for 
hexamethylbenzene in PS in the temperature range roughly $150-300 K$ 
\cite{RosslerEtAl90a}. 
Even if TEMPO and hexamethylbenzene have similar sizes their shape 
is rather different, then it is quite reassuring to note that the 
rotational motion of different {\it small} probes 
investigated by different techniques lead to comparable results.

Fig.\ref{FIG3} compares the PD model ( eq. \ref{eq:distrpower} , 
three adjustable parameters $\tau_{PD},\phi$ and $x$ ) with 
the ESR lineshape at both $190$ and $285 GHz$ of TEMPO in PS at $270 K$. 
Again, having fixed the jump angle, one adjusted the other two parameters. 
It is clearly seen that the agreement is quite improved, especially for 
small jump angles.  Fig. \ref{FIG4} provides the same comparison at 
$240 K$ for the case  $\phi = 20 {}^\circ$ which results in the best 
agreement.
{\bf From Figs.\ref{FIG3} and \ref{FIG4} it is seen that at both $240 K$ and 
$270 K$ the best-fit value of $\tau_{PD}$ is different for the two operating 
frequencies $190 GHz$ and $285 GHz$ with $\tau _{PD}( 190 GHz ) 
< \tau _{PD} ( 285 GHz )$. The uncertainty on $\tau_{PD}$ is roughly less than 
$20 \%$. Constraining $\tau _{PD}$ to having the same value at $190 GHz$ and 
$285 GHz$in the fit procedure  results in larger deviations between the 
experimental lineshape and the numerical results.  The decrease of $\tau_{PD}$
with the operating ESR frequency has been noted by us also at frequencies 
lower than $190 GHz$ and will be discussed elsewhere 
\cite{jcpfuture}. Without going into details, 
it has to be ascribed to the relation between the Larmor period
( roughly the inverse of the operating frequency  ) and the 
sensitivity of the ESR spectroscopy to the fast rotational dynamics.
}

Table \ref{Table1} summarizes the best-fit results of the PD model ( $\phi = 20 
{}^\circ$ ) for both $190$ and $285 GHz$. 
The width of the barrier-height distribution of TEMPO in PS is in the 
range $ 470 K \le \overline{E}/k \le 705 K$. One may wonder if 
this range corresponds to correlation times which are effectively accessed 
by HF-ESR in that the longest correlation time of TEMPO 
which may be measured is $\tau_{max} \cong 100 ns$. To this 
aim, one estimates 
the maximum energy barrier which TEMPO may overcome leading to 
appreciable motional narrowing effects in the lineshape as
$E_{max} \cong kT \; ln (\tau_{max}/\tau _{PD})$. In the temperature 
range $240K - 270K$ we get $E_{max}/k \cong 1500-1800 K $ which is 
fairly larger than $\overline{E}$, see Table \ref{Table1}. This 
confirms that the overall ESR lineshape, expressed by the superposition 
given by eq.\ref{eq:superpos}, is mostly due to components which are 
motionally narrowed to some or large extent by the reorientation of 
TEMPO. Table \ref{Table1} shows that the width of the barrier-height 
distribution of TEMPO in PS increases on cooling. This points to better 
coupling of TEMPO and PS at $ T \cong 240 K$. However, we remind that at 
lower temperatures the width decreases ( see Fig.\ref{FIG1} ). This 
non-monotonic behavior, as well the increase of $\overline{E}$ with the 
frequency which is also apparent in Table  \ref{Table1}, will be 
discussed in detail elsewhere \cite{jcpfuture}.

The exponential distribution ( eq. \ref{eq:energydistribution} ) of the 
barrier-heights of PS was evidenced by both 
internal friction \cite{Topp}, Raman \cite{sokolov0} and light scattering 
\cite{sokolov} measurements. The studies converted the mechanical \cite{Topp} 
and the optical \cite{sokolov0,sokolov} susceptibilities 
$\chi''(\nu)$ by eq.\ref{eq:scaling} to get $g(E)$ and estimated its width as 
$\overline{E}_{IF}/k = 760 \pm 40 K $ ,  
$\overline{E}_{Raman}/k = 530 \pm 60 K $
and $\overline{E}_{LS}/k = 530 \pm 40 K$, respectively.
It was concluded that, in spite of the different estimates of 
$\overline{E}$, the corresponding three patterns of $g(E)$ are 
in good agreement ( see ref.\cite{sokolov} and especially fig.4 ). 
Table \ref{Table1} shows that the width of the barrier-height 
distribution of TEMPO is in the range of the estimates on the width of 
the PS distribution provided by the above studies.

{\bf
One final remark concerns $\tau_{PD}$, i.e. the time scale $\tau_{0} $ 
of the activated trapping of TEMPO, eq. \ref{eq:Arrhenius}. The best-fit values 
of $\tau_{0}$ are about  $120-250 ps$ (~see captions of Figs. \ref{FIG3}, 
\ref{FIG4} ). Previous studies about the rotational dynamics of TEMPO in several 
glassy polymers were carried out by the customary X-band ESR spectroscopy at 
$9 GHz$ and analysed in terms of the SCT model \cite{Tormala1980,Cameron1982}. 
They yielded  $\tau_{0} \cong 10^{2}-10^{3} ps$ . 
In particular, for TEMPO in glassy PS it was found $\tau_{0} \cong 
10^{2} ps$ \cite{Tormala1980}. 
}

\section{Conclusions}
\label{conclusions}
The rotational motion of the guest molecule TEMPO in PS was investigated 
by using HF-ESR at two different Larmor frequencies ( $190$ and 
$285 \; GHz$ ). The use of a single correlation time ( SCT model ) was found 
to be inadequate to fit the ESR lineshape. Limited improvement was reached 
by considering a gaussian distribution of energy-barriers for TEMPO ( LGD 
model )  even if the best-fit width compared rather well with the one of 
molecular probes with similar size in PS as measured by NMR studies 
\cite{RosslerEtAl90a}. Assuming the exponential shape of the 
energy-barriers distribution for the spin probe (~PD model ) led to much better 
agreement at both $240\;K$ and $270\;K$. The same shape was also 
evidenced by other studies \cite{sokolov0,sokolov,Topp} on the distribution of 
the energy-barriers of PS which reported considerably different estimates of the 
width $\overline{E}$ (~$> 40 \%$~). The width of the barrier-height distribution 
of TEMPO was found in the range of the previous estimates of 
$\overline{E}$ for PS.

{ \bf The evidence that the deep structure of the energy landscape of PS
exhibits the exponential shape of the energy-barriers distribution agrees 
with results from extreme-value statistics \cite{Bouchaud2} and the trap 
model by Bouchaud and coworkers \cite{Bouchaud1,RinnEtAl2001}  }. 

\ack
Emilia La Nave and Francesco Sciortino are warmly thanked for useful discussions.

\Tables

\begin{table}
\caption{The width of the exponential distribution 
of barrier-heights $\overline{E}$ of TEMPO in PS at 
$T = 240K$ and $270 K$ as provided by the best-fit of the 
HF-ESR lineshapes at $190$ and $285 GHz$ by using the PD model with 
jump angle $\phi =20{}^\circ$. Previous measurements of the width of 
the distribution of barrier-heights of PS by internal 
friction \cite{Topp}, Raman \cite{sokolov0} and light scattering \cite{sokolov} 
yield 
$\overline{E}_{IF}/k = 760 \pm 40 K $ ,
$\overline{E}_{Raman}/k = 530 \pm 60 K $ and 
$\overline{E}_{LS}/k = 530 \pm 40 K$, respectively.
\label{Table1}
} 

\begin{indented}
\lineup
\item[]\begin{tabular}{@{}*{3}{c}}
\br                              
$\0\0T(K)$&$f(GHz)$&$\0\overline{E}/k ( K)$\cr 
\mr
\0\0$240$&$190$&\0$600 \pm 36$\cr
\0\0$240$&$285$&\0$705 \pm 42$\cr
\0\0$270$&$190$&\0$470 \pm 28$\cr
\0\0$270$&$285$&\0$540 \pm 32$\cr
\br
\end{tabular}
\end{indented}
\end{table}

\clearpage

\Figures

\Figure{\label{FIG1} The ESR line shapes at $190 GHz$ 
(left ) and $285 GHz$ ( right ) of TEMPO in PS at $50K$ . 
Magnetic parameters are: 
$g_{x} = 2.00994\pm 3\cdot 10^{-5}$,
$g_{y} = 2.00628\pm 3\cdot 10^{-5}$,
$g_{z} = 2.00212\pm 3\cdot 10^{-5}$,
$A_{x}(mT) = 0.62\pm 0.02$,
$A_{y}(mT) = 0.70\pm 0.02$,
$A_{z}(mT) = 3.40\pm 0.02$
The superimposed dashed lines are best-fits according to the SCT 
model 
eq. \ref{eq:deltadistr} with $\tau_{SCT} =25 ns$ ( $190 
GHz$ ) and $\tau_{SCT} = 19 ns$ ( $285 GHz$ ). Jump angle $\phi 
=60{}^\circ$.
The theoretical lineshapes were convoluted by a gaussian with width 
$w= 0.15 mT$ to account for the inhomogeneous broadening.
}

\Figure{\label{FIG2} Comparison of the EPR line shape at $190 GHz$ of 
TEMPO 
in PS at $270K$ with SCT and LGD models.
Magnetic parameters and gaussian width as in Fig.\ref{FIG1}.
Left: Best fits according to the SCT model , eq. \ref{eq:deltadistr}, 
with different jump angles.  The best-fit parameters are:
$\phi =5{}^\circ$ , $\tau_{SCT} =4.2 ns$; 
$\phi =20{}^\circ$ , $\tau_{SCT} =4.16 ns$; 
$\phi =60{}^\circ$ , $\tau_{SCT} =3.75 ns$; 
$\phi =90{}^\circ$ , $\tau_{SCT} =3.16 ns$. 
 Right:
Best fits according to the LGD model, eq.\ref{eq:distrloggauss}, 
with different jump angles. 
The best-fit value of the width was found to be independent of the 
jump angles 
and set to $\sigma =1.0$. The other best-fit parameters are:
$\phi =5{}^\circ$ , $\tau_{LGD} =3.6 ns$; 
$\phi =20{}^\circ$ , $\tau_{LGD} =3.6 ns$; 
$\phi =60{}^\circ$ , $\tau_{LGD} =1.8 ns$; 
$\phi =90{}^\circ$ , $\tau_{LGD} =1.6 ns$. 
}
\Figure{\label{FIG3} Best fits of the EPR line shape at $190 GHz$ ( 
left ) and $285 GHz$ ( right ) of TEMPO in PS at $270K$ according 
to the PD model, eq.\ref{eq:distrpower}, and different jump angles 
$\phi$. Magnetic parameters and gaussian width as in Fig.\ref{FIG1}. 
Best-fit parameters as follows. Left: 
$\phi =5{}^\circ ,  x=0.6 , \tau_{PD} =0.24 ns$; 
$\phi =20{}^\circ,  x=0.575 , \tau_{PD} =0.225 ns$; 
$\phi =60{}^\circ , x=0.6 ,\tau_{PD} =0.25 ns$; 
$\phi =90{}^\circ , x=0.63 ,\tau_{PD} =0.25 ns$. 
Right: 
$\phi =5{}^\circ ,  x=0.52 , \tau_{PD} =0.13 ns$; 
$\phi =20{}^\circ,  x=0.5 , \tau_{PD} =0.13 ns$; 
$\phi =60{}^\circ , x=0.5 ,\tau_{PD} =0.15 ns$; 
$\phi =90{}^\circ , x=0.55 ,\tau_{PD} =0.15 ns$. 
}
\Figure{\label{FIG4} 
The ESR line shapes at $T = 240K$ and
frequencies $190 GHz$ (panel a ) and $285 GHz$ (panel b).
The dotted superimposed curves are numerical simulations by using the
PD model ( eq.\ref{eq:distrpower} )
with $x=0.4 , \tau_{PD} =0.25 ns$ (panel a );
$x=0.34 , \tau_{PD} =0.12 ns$ (panel b ). In both cases
the best-fit value of the jump angle is $\phi =20{}^\circ$ . 
Magnetic parameters 
and gaussian width as in Fig.\ref{FIG1}.
}

\clearpage

\pagestyle{empty}

\begin{figure}
\begin{center}
\includegraphics[width=12cm]{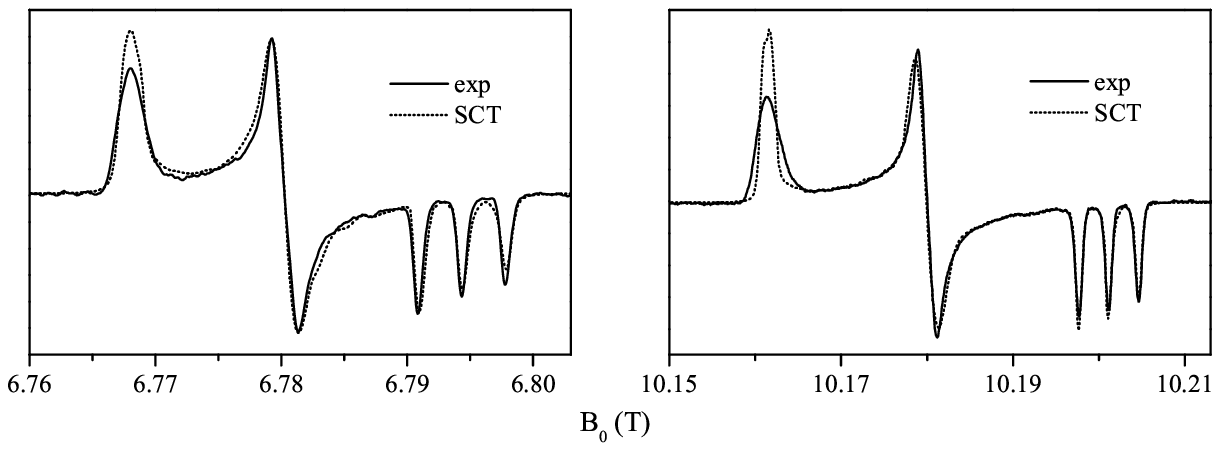}
\\
\vskip 1cm {\Large {\bf FIGURE 1}}
\end{center}
\end{figure}
\clearpage

\begin{figure}
\begin{center}
\includegraphics[width=12cm]{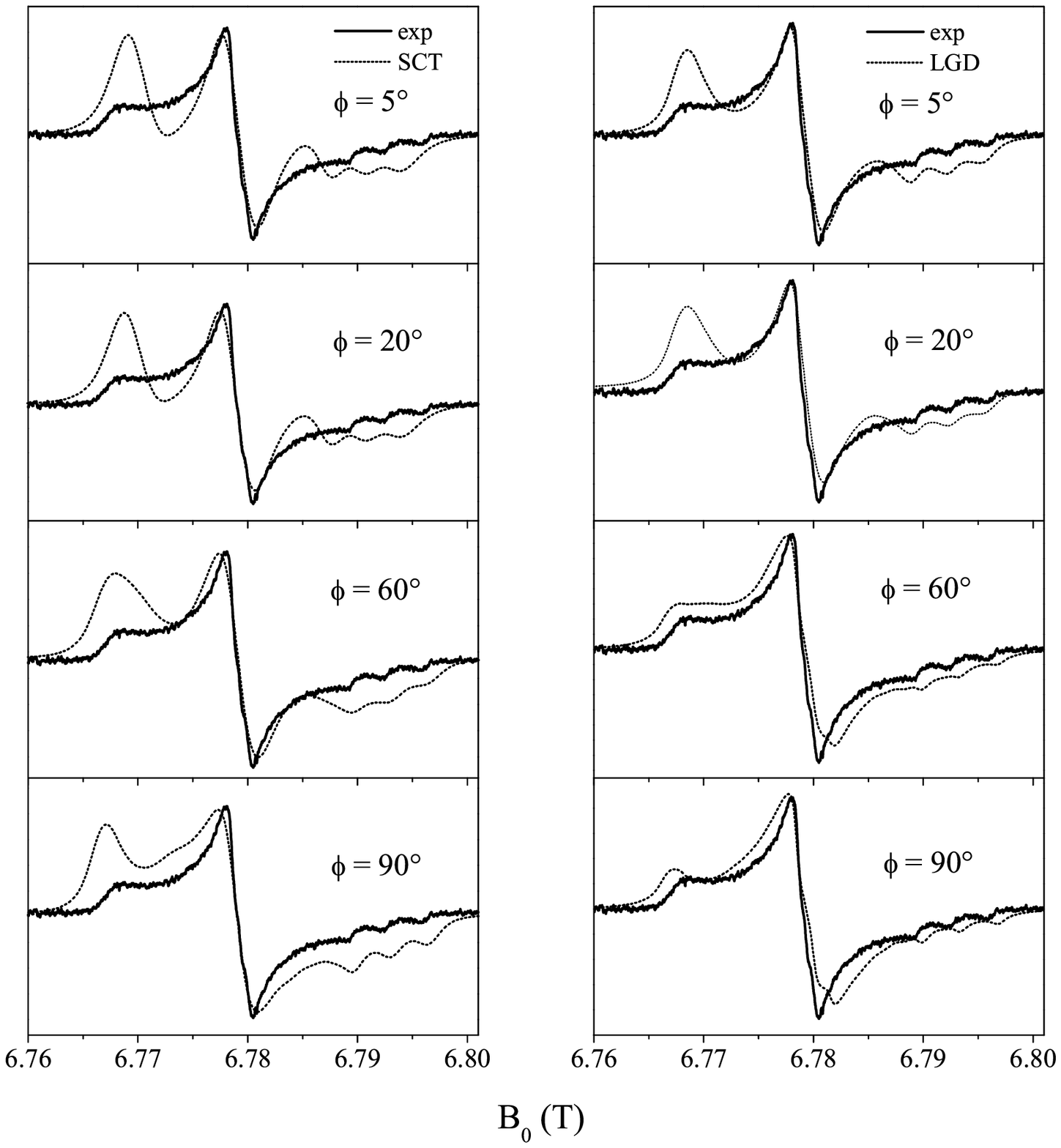}
\\
\vskip 1cm {\Large {\bf FIGURE 2}}
\end{center}
\end{figure}
\clearpage

\begin{figure}
\begin{center}
\includegraphics[width=12cm]{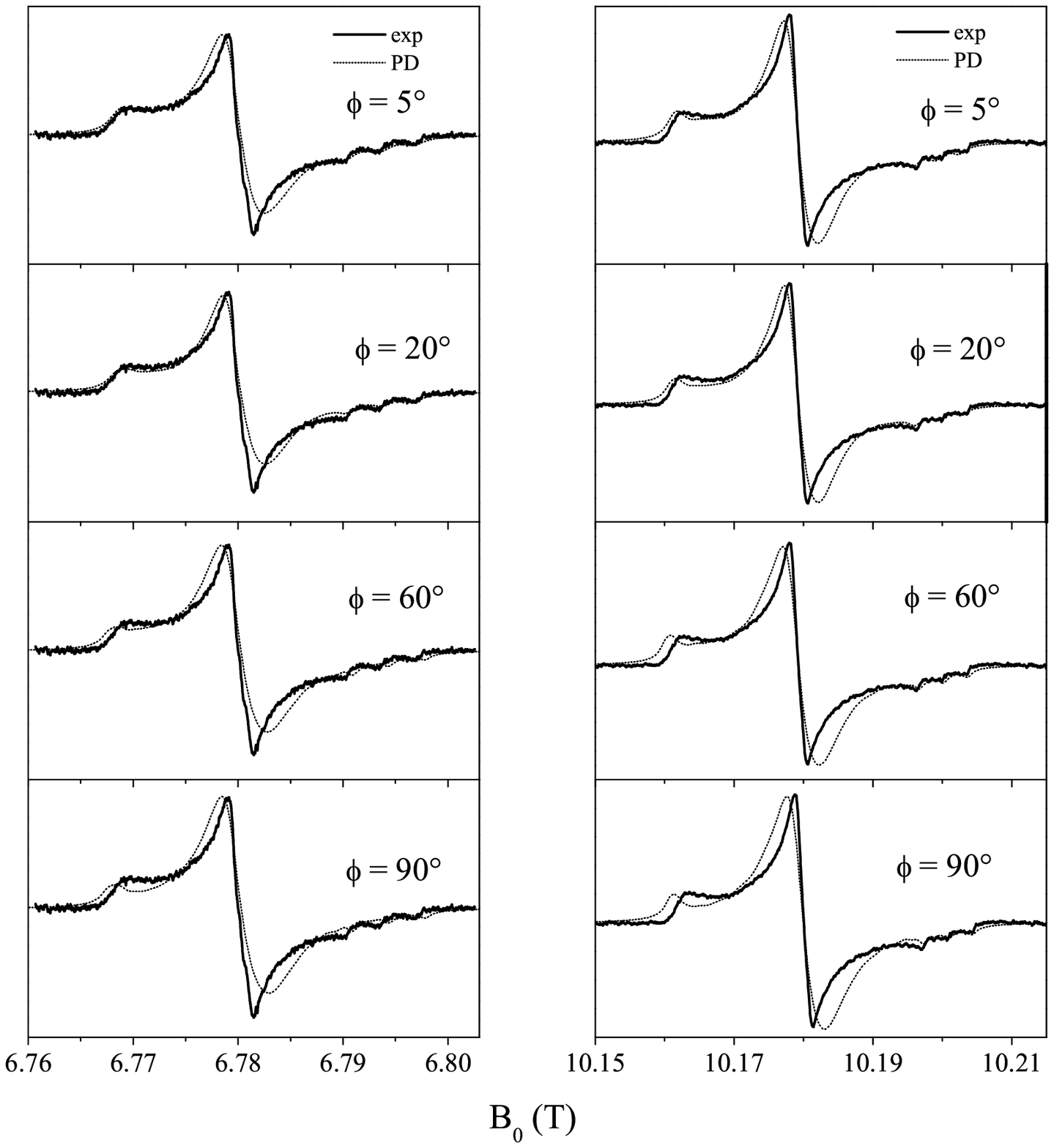}
\\
\vskip 1cm {\Large {\bf FIGURE 3}}
\end{center}
\end{figure}
\clearpage

\begin{figure}
\begin{center}
\includegraphics[width=8cm]{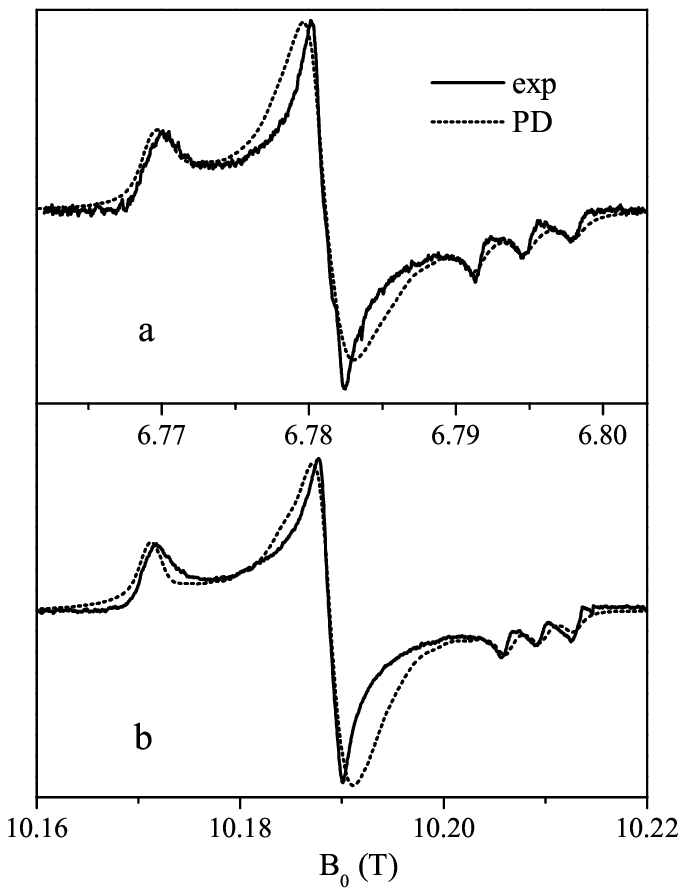}
\\
\vskip 1cm {\Large {\bf FIGURE 4}}
\end{center}
\end{figure}
\clearpage


\section*{References} 

\begin{thebibliography}{99}

\bibitem{AngEtAl00} Angell C A, Ngai K L, McKenna G B,
McMillan P F, Martin S W 2000 {\it J. Appl. Phys.} {\bf 88}, 3113.
\bibitem{review} B\"{o}hmer R, Diezemann G, Hinze G, R\"{o}ssler 
E 2001 
{\it Prog.Nucl.Mag.Reson.} {\bf 39}, 191.
\bibitem{Angell98} Angell C A 1998 {\it Nature (London)} {\bf 393}, 521.
\bibitem{Wu90} Wu L 1991 {\it Phys.Rev.B} {\bf 43}, 9906.
\bibitem{WuNagel90} Wu L, Nagel S R 1992 {\it Phys.Rev.B} {\bf 46}, 
11198.
\bibitem{qi} Qi F, B\"{o}hmer R, Sillescu H 2001 {\it 
Phys.Chem.Chem.Phis.} 
{\bf 3}, 4022.
\bibitem{JohGold70} Johari G P, Goldstein M 1970 {\it 
J.Chem.Phys.} {\bf 53}, 2372.
\bibitem{McCrumetal67} McCrum N G, Read B E , Williams G 1967 
{\it Anelastic and Dielectric Effects in Polymeric Solids}
( New York: Wiley )
\bibitem{rosslervogel1} Vogel M, R\"{o}ssler E 2001 {\it 
J.Chem.Phys.} {\bf 114}, 5802.
\bibitem{rosslervogel2} Vogel M, R\"{o}ssler E 2001 {\it 
J.Chem.Phys.} {\bf 115}, 10883.
\bibitem{sokolov0} Sokolov A P, Novikov V N, Strube B 1997 
{\it Europhys.Lett.} {\bf 38}, 49.
\bibitem{sokolov} Surovtsev N V, Wiedersich J A H , Novikov V N , 
R\"{o}ssler E , Sokolov A P 1998 {\it  Phys.Rev.B } {\bf 58}, 14888.
\bibitem{Topp} Topp K A , Cahill D G 1996 {\it Z.Phys.B} {\bf 101}, 
235.
\bibitem{GorhametAl2004} Gorham N T, Woodward R C, St.Pierre T G, 
Stamps R L, Walker M J, Greig D, Matthew J A D  2004 
{\it J.Apll.Phys.} {\bf 95}, 6983.
\bibitem{Derrida81} Derrida B 1981 {\it Phys.Rev.B} {\bf 24}, 2613.
\bibitem{Scher} Scher H, Montroll E W 1975 {\it Phys.Rev.B} {\bf 
12}, 2455.
\bibitem{Bassler87} B\"{a}ssler H 1987 {\it Phys.Rev.Lett.} {\bf 
58}, 767.
\bibitem{Odagaki95} Odagaki T 1995 {\it Phys.Rev.Lett.} {\bf 75}, 
3701.
\bibitem{Bouchaud1} Monthus C, Bouchaud J -P 1996 {\it J.Phys.A: 
Math.Gen.} {\bf 29}, 3847.
\bibitem{Bouchaud2} Bouchaud J -P, Mezard M 1997 {\it J.Phys.A: 
Math.Gen.} {\bf 30}, 7997.
\bibitem{RinnEtAl2001} Rinn B, Maass P, Bouchaud J-P 2001 
{\it Phys.Rev.B} {\bf 64}, 104417.
\bibitem{SchonSibani2000} Sch\"{o}n J C, Sibani P 2000 
{\it Europhys.Lett.} {\bf 49}, 196.
\bibitem{Bouchaud3} Denny R A, Reichman D R, Bouchaud J -P 2003 
{\it Phys.Rev.Lett.} {\bf 90}, 025503.
\bibitem{RosslerEtAl90a} R\"{o}ssler E, Taupitz M, Vieth H M 1990 
{\it J.Phys.Chem.} {\bf 94}, 6879.
\bibitem{Bordewijk} B\"{o}ttcher C J F, Bordewijk P 1978 
{\it Theory of Electric Polarization} ( Amsterdam: Elsevier).
\bibitem{RosslerEtAl90b} R\"{o}ssler E, Taupitz M, B\"{o}rner K, 
Schulz M, Vieth H M 1990 {\it J.Chem.Phys.} {\bf 92}, 5847.
\bibitem{AndrEtAl96} Andreozzi L, Cianflone F, Donati C, Leporini 
D 1996 {\it J.Phys.: Condens.Matter } {\bf 8}, 3795.
\bibitem{FaetEtAl99} Faetti M, Giordano M, Leporini D, Pardi L
1999 {\it Macromolecules } {\bf 32}, 1876.
\bibitem{BarbEtAl04} Barbieri A, Gorini G, Leporini D 2004 
{\it Phys. Rev. E} {\bf 69}, 061509.
\bibitem{ChristoffAtvars99} Christoff M, Atvars T D Z 
1999 {\it Macromolecules } {\bf 32}, 6093.
\bibitem{lebedev} Ondar M A ,Grinberg O Y ,Oranskii L G , 
Kurochkin V I , Lebedev Y. L. 1981 {\it J. Struct. Chem.} {\bf 22}, 626.
\bibitem{freed2} Budil D E , Earle K A , Freed J H 1993
{\it J. Phys. Chem.} {\bf 97}, 1294.
\bibitem{esrgunnar} Schweiger A , Jeschke G 2001 {\it Principles of 
Pulse Electron Paramagnetic Resonance} (Oxford: Oxford University Press ).
\bibitem{MartinelliEtAl2003} Martinelli M, Massa C A, Pardi L A, 
Bercu V, Popescu F F 2003 {\it Phys. Rev. B} {\bf 67}, 014425.
\bibitem{LiangFreed99a} Liang Z C, Freed J H 1999 
{\it J. Phys. Chem. B} {\bf 103}, 6384.
\bibitem{LiangEtAl99b} Liang Z C, Freed J H, Keyes R S, Bobst A M
2000 {\it J. Phys. Chem. B} {\bf 104}, 5373.
\bibitem{BorbatEtAl2001} Borbat P P, Costa-Filho A J, Earle K A, 
Moscicki J K, Freed J H 2001 {\it Science}, {\bf 291}, 266.
\bibitem{FuchsEtAl2002} Fuchs M R, Schleicher E, Schnegg A, Kay 
C W M, Torring J T, Bittl R, Bacher A, Richter G, Mobius K, Weber S 
2002 {\it J. Phys. Chem. B} {\bf 106}, 8885.
\bibitem{SchaferEtAl2003} Schafer K O, Bittl R, Lendzian F, 
Barynin V, Weyhermuller T, Wieghardt K, Lubitz W 2003
{\it J. Phys. Chem. B} {\bf 107}, 1242.
\bibitem{PilarEtAl00} Pilar J, Labsky J, Marek A, Budil D E, 
Earle K A, Freed J H 2000 {\it Macromolecules } {\bf 32}, 4438.
\bibitem{LepEtAl02a} Leporini D, Zhu X X, Krause M , Jeschke G , 
Spiess H W 2002 {\it Macromolecules} {\bf 35}, 3977.
\bibitem{LepEtAl02b} Leporini D, Sch\"{a}dler V , Wiesner U , 
Spiess H W, Jeschke G 2002 {\it J. of Non-Crystalline Solids} {\bf 
307Ð310}, 510.
\bibitem{LepEtAl03} Leporini D, Sch\"{a}dler V , Wiesner U , 
Spiess H W, Jeschke G 2003 {\it J.Chem.Phys.} {\bf 119}, 11829.
\bibitem{berliner}  Berliner L J edt. 1976 {\it Spin Labeling: 
Theory and Applications} ( New York: Academic Press); Berliner 
L J, Reuben J edts. 1989 {\it  Biological Magnetic Resonance}, 
Vol.8 (New York: Plenum).
\bibitem{Muus} Muus L T, Atkins P W edts. 1972 {\it Electron Spin 
Relaxation in Liquids} ( New York: Plenum ).
\bibitem{SaalmuellerEtAl96} Saalmueller J W, Long H W , Volkmer T, 
Wiesner U, Maresch G G, Spiess H W 1996 {\it J.Polymer Science:part B, 
Polymer Physics} {\bf 34}, 1093.
\bibitem{AnninoEtAl00} Annino G, Cassettari M , Fittipaldi M , 
Lenci L, Longo I , Martinelli M , Massa C A , Pardi L A 2000
{\it Appl. Magn. Reson.}  {\bf 19} ,495.
\bibitem{jcpfuture} Bercu V, Pardi L A, Massa C A , Martinelli M, 
Leporini D to be submitted.
\bibitem{Tormala1980} T\"{o}rm\"{a}l\"{a} P in {\it Molecular Motion in 
Polymers by ESR} Boyer R F, Keinath S E  edts. 1980 ( Chur: Harwood )
\bibitem{Cameron1982} Cameron G G 1982 {\it Pure and Appl.Chem.} {\bf 54}, 
483.

\end{thebibliography}
\end{document}